\documentclass{Rinton-P10x7}

\begin{document}

\title{Theoretical issues in spin-based quantum dot
quantum computation}

\author{Xuedong Hu and S. Das Sarma}

\address{Department of Physics, University of
Maryland, College Park, MD 20742-4111 \\
E-mail: xdhu@physics.umd.edu, dassarma@physics.umd.edu
}

\maketitle

\abstracts{
We review our recent work addressing various theoretical issues in
spin-based quantum dot quantum computation and quantum information
processing.  In particular, we summarize our calculation of
electron exchange interaction in two-electron double quantum dots
and multi-electron double dots, and discuss the physical implication
of our results.  We also discuss possible errors
and how they can be corrected in spin-based quantum dot quantum
computation.  We critically assess the constraints and conditions
required for building spin-based solid state quantum dot quantum
computers.   
}

\section{Introduction}
\label{section1}

It has long been pointed out that quantum mechanics 
may provide great advantages over classical physics in physical
computation.\cite{Deutsch,Feynman}  However, the expeditious growth
of research on quantum computation\cite{review} started after the
advent of Shor's factorization algorithm\cite{Shor1} and quantum error
correction schemes.\cite{Shor2,Steane}  Among the many hardwares that
were proposed are the ones based on electron
spins.\cite{DiVincenzo,LD,Vrijen,Imam,special}  Obviously, electron
spin provides a perfect candidate for quantum bits
(qubits) as the electron spin sub-Hilbert-space is generally
well-defined and its decoherence relatively slow.  In addition,
electron spin can be manipulated by external magnetic fields and by
interacting with other spins through exchange or Heisenberg coupling.

Our study of electron-spin-based quantum computers (QC) has focused
on the scheme using single electrons trapped in quantum
dots.\cite{LD,BLD,HD}  Here the qubit is the spin degrees of freedom
of a single electron trapped in a gated semiconductor (GaAs)
horizontal quantum dot (QD). Single-qubit operations can be achieved
by applying an external magnetic field, while two-qubit operations
can be built upon exchange interaction between electrons in
neighboring quantum dots when the inter-dot potential barrier is
lowered and electron orbital wavefunction is allowed to mix.  Final
measurement in this scheme requires single spin detection, for which
various proposals such as spin filters\cite{DPD1} have been
suggested.  

A working quantum computer needs to satisfy many stringent
conditions.  The main requirements include initialization,
well-defined Hilbert space, weak decoherence, precise coherent
manipulation, unitary evolution, and efficient quantum
measurement.\cite{DPD3}  In the case of spin-based quantum dot
quantum computer, electron spin can be initialized by lowering
the sample temperature, and applying an external magnetic field to
differentiate the spin up and down directions.  Because of the
three-dimensional confinement and the fact that GaAs conduction
band is mainly formed from S atomic orbitals, the trapped electron
has very weak spin-orbit coupling and therefore its spin has a small
decoherence rate.  The spin Hilbert space of a single quantum dot
with one electron is well-defined as long as the Zeeman splitting
between the two spin levels is much smaller than the single electron
excitation energy associated with the quantum dot confinement
potential, which requires a sufficiently small quantum dot.  The
Hilbert space structure is not as clear cut when two quantum dots are
brought together for two-qubit operations. The exploration of this
part of the spin Hilbert space is one major focus of our recent
study, and we will review some of our results in the following
section.  Coherent control of electrons in mesoscopic nanostructures
has been under intensive study by many groups for
sometime.\cite{Ashoori,Waugh,Liver,Blick,Oost,Kotlyar,Banin,Fuji,Ji} 
To reliably manipulate individual electrons and their spins, more
precise controlling techniques have to be developed, and various
noise effects such as those from offset charge oscillations and other
volatile sources have to be eliminated or suppressed.  Single spin
detection is another difficult proposition for experimentalists. 
Currently available SQUID (superconducting quantum interference
device) can detect a single Bohr magneton but requires a long
detection time (in the order of minutes), while proposed spin
valve techniques have not yet succeeded in detecting the spin of a
single electron.  Converting the spin signal to a charge signal and
then using single electron transistor based charge measurement
techniques to measure the electron spin seems one promising
experimental method, but whether this technique (or any other spin
measurement technique) can be implemented in microsecond type time
scales (the typical low temperature spin coherence time is expected
to be microseconds in a practical QC) remains to be seen. 

In Section~\ref{section2} we summarize our work on the theoretical
exploration  of double dot electron spin Hilbert space, including
both two electrons and six electrons in a double dot.  In
Section~\ref{section3} we discuss possible errors in a quantum dot
quantum computer.

\section{Electron Spin-based Quantum Computation}
\label{section2}

\begin{figure}[t]
\epsfxsize=30pc 
\epsfbox{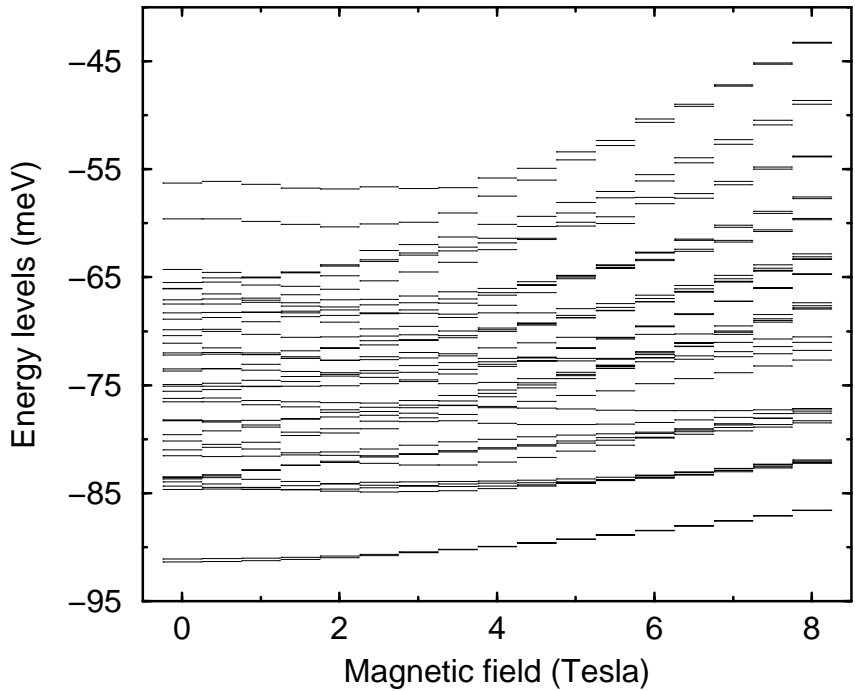} 
\caption[]{A sample energy spectra of a two-electron double quantum
dot in GaAs\cite{HD}.  Here we show the magnetic field dependence of
the two-electron energy spectra obtained in a molecular orbital
calculation where both S and P single-electron Fock-Darwin orbitals
are used.  The dot radius is 20 nm, the inter-dot distance is 30 nm,
and the central barrier $V_b$ is 30 meV, corresponding to an actual
barrier height of 9.61 meV.  
\label{fig1}}
\end{figure}

We have studied a double quantum dot (each has one trapped electron)
as the basic elementary gate for a quantum computer.\cite{HD}  This
can be considered an artificial double dot quantum molecule with two
electrons, analogous to the H$_2$ molecule with one electron on each
dot.  In the first order approximation, a Heisenberg exchange
Hamiltonian can be written to describe the dynamics of the two
electrons trapped in the double dot.  However, to establish the
operability of such an exchange-based quantum computer, we need to
clearly delineate the relation between the two-spin Hilbert space and
the complete two-electron Hilbert space.  For this purpose, we use a
molecular orbital calculation to determine the excitation spectrum of
two electrons in two horizontally coupled GaAs quantum dots, and study
its dependence on an external magnetic field.  In order to provide a
more accurate description of the low lying excited states, we include
the first excited single particle orbital states (P orbitals of the
Fock-Darwin states,\cite{Jacak} which are the eigenstate sequence of a
two-dimensional electron in an external perpendicular magnetic
field).  The resulting spectrum (with an example shown in
Fig.~\ref{fig1}) confirms the belief that the exchange spin
Hamiltonian is sufficient to describe the low energy dynamics of the
two electrons.  

As shown in Fig.~\ref{fig1}, the lowest energy states (one singlet
and one triplet) are well-separated from the rest of the excited
states.  Indeed, the higher energy excited states are separated
from the lower energy subspace by the smaller of the single particle
excitation energy and the on-site Coulomb repulsion energy (the
so-called Coulomb blockade energy) in a single dot.   This energy
separation is at least 5 meV in our cases, much larger than the GaAs
electron Zeeman splitting ($\approx 2.55 \times 10^{-2}$ meV/Tesla)
and the energy separation between the ground singlet and triplet
states (in the order of 0.1 meV).  If the adiabatic condition is
satisfied when the exchange gate is turned on, the ground states are
essentially isolated from the higher energy excited states.  The
orbital composition of the ground singlet and triplet states are
basically $\psi_{L0}$ and $\psi_{R0}$, the lowest Fock-Darwin
orbitals, which are also the orbital states of the electrons when
they are in separate quantum dots.  Therefore, throughout the
duration of the exchange gate (or the lowering of the inter-dot
potential barrier), the electron orbital degrees of freedom are
frozen, and a spin Hamiltonian such as the Heisenberg exchange
Hamiltonian in this case can describe the complete two-electron
adiabatic dynamics. 

The low-energy spin subspace of the complete two-electron Hilbert
space consists of a singlet and three triplet states (whose
degeneracy can be lifted by an external magnetic field).  The
splitting between the singlet and the unpolarized triplet state is
the key parameter that determines the speed of two-qubit operations
such as an exact swap of the spin states and controlled NOT (CNOT).
In essence, the exchange interaction generates a phase difference
between the singlet and the unpolarized triplet states (but not the
relative weight) in a general two-spin state, thus changing the
density matrix of each individual spin.  Our numerical results show
the singlet-triplet splitting J (also called exchange constant in
this context) to be in the range of 0.01 to 1 meV as we vary the
inter-dot distance and the size of the individual dots.  Such a
magnitude of J corresponds to reasonably short operating times (in
the order of 0.1 to 10 ns after the adiabatic condition is taken into
account) compared to the estimated electron spin coherence time
(possibly of the order of microseconds at 50-100 mK temperature) in
semiconductor heterostructures.\cite{DFHZ,Awschalom}  Note that low
temperature ($T$) QD-QC operation is essential in this context: first,
$T$ must be low to obtain a long spin coherence time; and second, $k_B
T$ must be low to satisfy $\Delta \gg k_B T$ condition, where
$\Delta$ is the separation between the ground and the higher excited
orbital levels in the single quantum dot system ($\Delta > 1$
meV in small dots), to maintain the integrity of the spin sub-Hilbert
space necessary for quantum computation.  Thus, the operation
(or the manipulation) time for various gates and magnetic field
pulses in QD-QC architecture is constrained from below by the
adiabaticity requirement (i.e. it cannot be too fast so that
adiabaticity is maintained) and from above by the spin coherence
requirement (i.e. the operation must be much faster than the spin
coherence time).  One must realize that in addition to these
stringent operational constraints there is also the severe (and as
yet experimentally unsolved) constraints of measuring the final spin
state (``read out'') which cannot be too slow for successful QD-QC
operations.

Our molecular orbital calculation is based on the effective mass
approximation in a single envelope-function approach.  Due to
the sizable band gap in GaAs and the relatively small inter-band
coupling, our single envelope function calculation is valid with quite
high accuracy. Indeed, we estimated\cite{HD} that the effect of the
neglected  bands and inter-conduction band coupling amounts to about
1\% correction to the single particle wavefunctions.  To clearly
quantify this correction one would have to employ a more sophisticated
multi-band calculation or drop the envelope function approach
entirely and adopt an atomistic approach, which would automatically
include all valence band effects.  However, it is unclear, in
principle, whether such a first principle type band theory approach
would be a meaningful calculational tool for assessing QD-QC
operations since the important energy scales (e.g. J) in the problem
are 1 meV or less, and first principle band calculations cannot
achieve 1 meV type accuracies.

Experimentally, it is very difficult to trap one and only one
electron in a gated two-dimensional GaAs quantum dot.  Thus it has
been speculated that perhaps the requirement of single trapped
electron per quantum dot as the fundamental qubit may be
relaxed in a QD-QC, and that an odd number of
electrons in a QD can provide an effective spin-1/2 system
as a qubit.  To explore this possibility, we have recently done a
multi-electron calculation where we have three electrons in one
quantum dot, thus six electrons in a double dot.\cite{HD2}  We first
carry out a Hartree Fock mean field calculation to produce an
orthonormal set of single-particle basis states, then employ a
restricted configuration interaction (CI) approach to calculate the
energy spectrum.  Our results show that the low energy dynamics in the
multi-electron case is generally much more complicated than the
two-electron case because of the single-particle energy level
degeneracies in the higher excited Fock Darwin states.  This
complexity can be alleviated by applying external magnetic fields or
deforming the quantum dots, which lift the level degeneracies.  Thus
we show in Ref.~\cite{HD2} that subject to some constraints
spin-based QD-QC architectures are feasible, at least in principle,
in multielectron quantum dot arrays.

\section{Error correction in quantum dot quantum computer}
\label{section3}

To evaluate the feasibility of a quantum computer architecture, it is
only one of the many steps to study the operating Hilbert space
structure.  Having a well defined sub-Hilbert space, as in
Fig.~\ref{fig1}, is a necessary but by no means a sufficient
condition for QC operation.  Another important step is to assess the
importance of all the possible errors (and how to do the relevant
error correction) in this architecture.  For spin-based quantum dot
quantum computer model, we have explored several kinds of operating
error (aside from spin decoherence and errors in single spin
measurement).  For example, we originally speculated that double
occupation in each of the double dot during a two-qubit operation
might result in a significant error, the rationale being that one
needs to keep tags on all the electrons as each represents a distinct
qubit.  We now believe that adiabatic operations of these two-qubit
gates should effectively erase this problem since  adiabaticity will
limit the two-electron dynamics within the low-energy two-spin
subspace as we discussed in the previous section.\cite{Ruskai}  We
have done simulations to study the adiabatic condition and the
results will be published elsewhere.  We no longer believe virtual
double occupancy to be a problem.

As the exchange coupling $J$ (the singlet-triplet splitting) is tuned
by changing external gate voltage in a QD-QC, thermal fluctuations 
(or any other types of fluctuations) in the gate voltage will
lead to fluctuations in $J$, thus causing phase errors in the
exchange-based swap gate which is crucial for two-qubit operations. 
We have estimated this error by assuming a simple thermal (white)
noise.\cite{HD}  More specifically, we assume $J=f(V)$ where 
$V$ is the gate voltage that controls the value of $J$.  Around
any particular value $V_0$, $J$ can be expressed as $J(V) =
J(V_0) + \left. f^{\prime}(V) \right|_{V_0} (V-V_0)$.  During a
swap gate between two quantum dots, the phase of the electronic 
spin wavefunction evolves as $\phi = \int_0^t J(\tau) d\tau /\hbar$.
Thus the fluctuation in the phase $\phi$ is 
\begin{eqnarray}
\langle \delta \phi^2 \rangle & = & \langle \phi^2 \rangle
- \langle \phi \rangle^2 \ = \ \frac{1}{\hbar^2} \int_0^t \int_0^t
\langle \delta  J(\tau_1) \ \delta J(\tau_2) \rangle d\tau_1 d\tau_2 
\nonumber \\
& \sim & \int_0^t \int_0^t 
\frac{[f^{\prime}(\bar{V})]^2}{\hbar^2}
\langle \delta V(\tau_1) \ \delta V(\tau_2) \rangle 
d\tau_1 d\tau_2 \,.
\end{eqnarray}
Using Nyquist theorem $\langle \delta V(\tau_1) \ \delta V(\tau_2)
\rangle = 4 R k_B T \delta(\tau_1-\tau_2)$,
we obtain the approximate expression for the phase fluctuation:
\begin{equation}
\langle \delta \phi^2 \rangle
\sim 4 R k_B T \alpha^2 t /\hbar^2 \,,
\end{equation}
where $\alpha$ is the upper bound of $|f^{\prime}(\bar{V})|$.
Assuming the swap gate is performed at 1 K
(since J is in the order of 0.1 meV $\sim$ 1 K, the
experimental temperature can only be lower than 1 K),
and the transmission line connecting the gate to the
surrounding bath of cryogenic temperature has an impedance
of 50 ohm, the rate for phase fluctuation $\langle \delta \phi^2
\rangle/t$ is about 3.2 MHz. The phase error accrued during a swap
gate is then about  0.06\%.  This is quite a small error which is
the same order of magnitude as the theoretical tolerance of the 
currently available quantum error correction codes.
To further lower this error rate, one can go to lower
experimental temperature and turn up $J$ more gently (which requires
longer time but produces smaller $\alpha$) in the QD-QC operation.

Another possible error in the two-qubit operations of the
architecture we studied is caused by inhomogeneous magnetic
fields.\cite{HDD}  Such a field may come from magnetic impurities or
unwanted currents away from the structure.  Magnetic field affects
both orbital and spin part of the electron wavefunction.  The orbital
effect is accounted for by adjusted exchange coupling J, while the
spin effect is accounted for through Zeeman coupling terms:
\begin{eqnarray}
H_s & = & J({\bf B}) {\bf S}_1 \cdot {\bf S}_2 + \gamma_1 S_{1z} 
+ \gamma_2 S_{2z} \,,
\label{eq:spin-Hamiltonian} 
\end{eqnarray}
where ${\bf S}_1$ and ${\bf S}_2$ refer to the spins of the two
electrons; $J({\bf B})$ is the exchange coupling (singlet-triplet
splitting); $\gamma_1$ and $\gamma_2$ are the effective strength of
the Zeeman coupling in the two quantum dots.  In an inhomogeneous
field, $\gamma_1 \neq \gamma_2$, so that the Zeeman terms do not
commute with the exchange term in the Hamiltonian 
(\ref{eq:spin-Hamiltonian}).  We have done a detailed analysis on
how to achieve swap with such a Hamiltonian, and found that there is
at the minimum an error proportional to the square of field
inhomogeneity in the swap.  For example, if the initial state of the
two electron spin is $|\phi(0)\rangle=|\!\uparrow \downarrow \rangle$,
the density matrix of the first spin after the optimal swap
is
\begin{equation}
\left. \rho_1 \right|_{e^{i\theta}=-1}
=\frac{1}{1+x^2}|\!\downarrow\rangle \langle \downarrow\!|
+ \frac{x^2}{1+x^2}|\!\uparrow\rangle \langle \uparrow\!| \,,
\label{eq:swap_error}
\end{equation}
where $x=\delta/(2J)=(\gamma_1-\gamma_2)/2J$.  In other words, 
the first spin can never exactly acquire the state ($|\!\downarrow 
\rangle$) of the second spin.  Its state will remain mixed and
the smallest error from an exact swap is $\frac{x^2}{1+x^2}$,
which needs to be corrected.  We have estimated\cite{HDD} that in
GaAs a Bohr magneton can lead to an error in the order of $10^{-6}$,
which is within the capability of currently available quantum
error correction schemes.

\section{Conclusion}

In this paper we have reviewed some of our recent results in our study
of electron-spin-based quantum dot quantum computation.  We discuss
various issues regarding the Hilbert space structure of a
double quantum dot artificial molecule.  We also describe possible
errors in a spin-based QC.  Our tentative conclusion is that building
a practical GaAs electron-spin-based QD-QC, while being certainly
possible in principle, will be extremely difficult in practice (even
assuming a working temperature of 100 mK or less) because of severe
constraints and limitations arising from the spin coherence time, the
adiabatic condition, the magnitude of exchange coupling, the error
correction requirement, and the unsolved experimental problem of a
reasonably fast (in less than a microsecond) measurement of a single
Bohr magneton.

\section*{Acknowledgments}
We thank financial support from ARDA, LPS, DARPA, and US-ONR.

\end{document}